\documentclass[prl,print£¬superscriptaddress,showpacs,onecolumn,showkeys]{revtex4}
\usepackage[dvips]{graphics,color}
\usepackage{amsfonts}
\usepackage{amsmath}
\usepackage{amssymb}
\usepackage{graphicx}
\usepackage{fancyhdr}

\setcitestyle{open={},close={}}

\input{tcilatex}
\begin{document}

\title{\textbf{MAXIMUM VIOLATION OF WIGNER INEQUALITY FOR TWO-SPIN ENTANGLED
STATES WITH PARALLEL AND ANTIPARALLEL POLARIZATIONS}}
\author{YAN GU, HAIFENG ZHANG, ZHIGANG SONG}
\affiliation{Institute of Theoretical Physics and Department of Physics, {State Key
Laboratory of Quantum Optics and Quantum Optics Devices, }Shanxi University,
Taiyuan, Shanxi 030006, China}
\author{J. -Q. LIANG}
\affiliation{{Institute of Theoretical Physics and Department of Physics, State Key
Laboratory of Quantum Optics and Quantum Optics Devices, Shanxi University,
Taiyuan, Shanxi 030006, China} \\
{*jqliang@sxu.edu.cn}}
\author{L. -F. WEI}
\affiliation{State Key Laboratory of Optoelectronic Materials and Technologies, School of
Physics and Engineering, Sun Yat-Sen University, Guangzhou 510275, China}
\affiliation{Quantum Optoelectronics Laboratory, School of Physics and Technology,
Southwest Jiaotong University, Chengdu 610031, China}
\received{4 March 2018}
\revised{24 May 2018}

\begin{abstract}
The experimental test of Bell's inequality is mainly focused on
Clauser-Horne-Shimony-Holt (CHSH) form, which provides a quantitative bound,
while little attention has been paid on the violation of Wigner inequality
(WI). Based on the spin coherent state quantum probability statistics we in
the present paper extend the WI and its violation to arbitrary two-spin
entangled states with antiparallel and parallel spin-polarizations. The
local part of density operator gives rise to the WI while the violation is a
direct result of non-local interference between two components of the
entangled states. The Wigner measuring outcome correlation denoted by $W$ is
always less than or at most equal to zero for the local realist model ($%
W_{lc}\leq 0$) regardless of the specific initial state. On the other hand
the violation of\ WI is characterized by any positive value of $W$, which
possesses a maximum violation bound $W_{\max }$ $=1/2$. We conclude that the
WI is equally convenient for the experimental test of violation by the
quantum entanglement.
\end{abstract}

\maketitle


%
%%%%%%%%%%%%%%%%%%%%%%%%%%%%%%%%%%%%%%%%%%%%%%%%%%%%%%%%%%%%%%%%%%%%

%\accepted{(Day Month Year)}
%\comby{(xxxxxxxxxx)}
\thispagestyle{fancy}

\part{\protect\large 1.\ Introduction}

The non-locality{\textsuperscript{\cite{1,2}}} as one of the most striking
characteristic of quantum mechanics does not have classical correspondence
within our intuition of space and time in the classical field theory.
Quantum entangled-state, which originally was introduced by
Einstein-Podolsky-Rosen to question the completeness of quantum mechanics,
has become a key concept of quantum information and computation.{%
\textsuperscript{\cite{3,4,5,6,7}}} From a two-spin entangled state proposed
by Bohm,{\textsuperscript{\cite{8}}} Bell proved a quantitative criteria
between the quantum and classical measuring-outcome correlations{%
\textsuperscript{\cite{9}}} known as Bell's inequality (BI). It was
established by means of classical statistics with the assumption of hidden
variable. The BI, which plays a fundamental role in quantum entanglement,
has attracted great attentions both theoretically and experimentally.{%
\textsuperscript{\cite{10,11,12,13,14}}}\textbf{\ }Bell nonlocality and
quantum entanglement in two-qubit spin model are also measured by use of
measurement induced disturbance and quantum discord.{\textsuperscript{%
\cite{15}}} The experimental evidence{\textsuperscript{%
\cite{13,14,15,16,17,18,19,20,21}}} confirming the violation of BI provides
an overwhelming superiority for the non-locality in quantum mechanics
against the proposition of local realism.{\textsuperscript{\cite{22,23}}} In
various modified forms of the BI, the Clauser-Horne-Shimony-Holt (CHSH)
inequality{\textsuperscript{\cite{24}}} is of particular interesting to the
experimental test, since it provides a quantitative bound of the
four-direction measuring-outcome correlation $P_{CHSH}^{lc}\leq 2$ for the
local realist theory. The inequality can be violated by the two-spin
entangled state with a maximum violation known as $P_{CHSH}^{\max }=2\sqrt{2}
$.

Recently a quantum mechanical framework was presented to formulate the
various forms of BI and their violation in a unified formalism.{%
\textsuperscript{\cite{25,26}}} The density operator of a bipartite
entangled-state is separated into the local and non-local (interference)
parts. The measuring outcome correlation is then evaluated by the quantum
probability statistics in the spin coherent-state base vectors along the
measuring directions.{\textsuperscript{\cite{27,28}}} The local part gives
rise to the local-realist correlation, which results in the BI, while the
non-local one is responsible for the violation. For the arbitrary high spins
a spin-parity effect in the violation of BI is found{\textsuperscript{%
\cite{25,26}}} as a result of Berry phase interference of spin coherent
states.

The Wigner inequality{\textsuperscript{\cite{29,30}} (WI)} is a simpler
form, in which the particle number probability of positive spin is measured.
We in the present work reformulate the WI and its violation for arbitrary
two-spin entangled states with both antiparallel and parallel polarizations.{%
\textsuperscript{\cite{31,32}}} The measuring outcome correlation is
evaluated by the spin coherent-state quantum probability statistics.
Although the WI is simple it attracts a little attention of experimenters.{%
\textsuperscript{\cite{33,34,35}}} The reason may be that it lacks a
quantitative bound, which is convenient for the experimental verification.
Following CHSH we propose a Wigner correlation $W$, which is\ less then or
at most equal to zero (equivalent to the WI) according to the local realism.
The maximum violation of WI is also found for the arbitrary two-spin
entangled states with both antiparallel and parallel polarizations.

\part{\protect\Large 2. Spin coherent-state quantum probability statistics
for measuring outcome correlation}

The original BI and the modified form CHSH inequality are derived based on
classical statistics with hidden variable assumption. In previous
publications{\textsuperscript{\cite{25,26}}} the Bell-type inequalities and
their violation are formulated in a unified manner by means of the spin
coherent-state quantum probability statistics. The density operator of an
entangled state for a bipartite system can be separated to the local (or
classical) and non-local (or quantum coherent) parts. The former part gives
rise to the local realist bound of measuring outcome correlation, namely the
BIs, while the latter part leads to the violation of the inequalities. We
begin with an arbitrary two-spin entangled state of antiparallel
polarization in the bases $\hat{\sigma}_{z}|\pm \rangle =\pm |\pm \rangle $
that%
\begin{equation}
|\psi \rangle =c_{1}|+,-\rangle +c_{2}|-,+\rangle ,  \label{1}
\end{equation}%
where the normalized coefficients can be generally parameterized as $%
c_{1}=e^{i\eta }\sin \xi $, $c_{2}=e^{-i\eta }\cos \xi $. We assume that two
spins are separated to a space-like distance when the entangled state is
prepared. The density-operator $\hat{\rho}$\ of entangled state can be
divided into two parts\ 
\begin{equation}
\hat{\rho}=\hat{\rho}_{lc}+\hat{\rho}_{nlc}.  \label{2}
\end{equation}%
The local part 
\begin{equation*}
\hat{\rho}_{lc}=\sin ^{2}\xi |+,-\rangle \left\langle +,-\right\vert +\cos
^{2}\xi |-,+\rangle \left\langle -,+\right\vert ,
\end{equation*}%
which is the classical two-particle probability-density operator, describes
the individual spin of the bipartite system separated remotely. While what
we called the non-local part%
\begin{equation*}
\hat{\rho}_{nlc}=\sin \xi \cos \xi \left( e^{2i\eta }|+,-\rangle
\left\langle -,+\right\vert +e^{-2i\eta }|-,+\rangle \left\langle
+,-\right\vert \right)
\end{equation*}%
is the quantum coherence density-operator between two remote spins.

\part{\protect\large 2.1 Spin measuring outcome correlation and violation of
BI}

The measurements of two spins are performed independently along two
arbitrary directions, say\textbf{\ }$\mathbf{a}$\textbf{\ }and\textbf{\ }$%
\mathbf{b}$. The measuring outcomes fall into the eigenvalues of projection
spin-operators\textbf{\ }$\hat{\sigma}\cdot \mathbf{a}$\textbf{\ }and\textbf{%
\ }$\hat{\sigma}\cdot \mathbf{b}$, i.e.\textbf{\ }%
\begin{equation*}
\hat{\sigma}\cdot \mathbf{a|}\pm \mathbf{a}\rangle =\pm \mathbf{|}\pm 
\mathbf{a}\rangle ,\quad \hat{\sigma}\cdot \mathbf{b|}\pm \mathbf{b}\rangle
=\pm \mathbf{|}\pm \mathbf{b}\rangle ,
\end{equation*}%
according to the quantum measurement theory. Solving the eigenvalue equation
for each direction denoted by $\mathbf{r}=\mathbf{a},\mathbf{b}$, we have
two orthogonal eigenstates given by 
\begin{align}
\left\vert +\mathbf{r}\right\rangle & =\cos \frac{\theta _{r}}{2}\left\vert
+\right\rangle +\sin \frac{\theta _{r}}{2}e^{i\phi _{r}}\left\vert
-\right\rangle ,  \notag \\
\left\vert -\mathbf{r}\right\rangle & =\sin \frac{\theta _{r}}{2}\left\vert
+\right\rangle -\cos \frac{\theta _{r}}{2}e^{i\phi _{r}}\left\vert
-\right\rangle .  \label{3}
\end{align}%
In the above solutions the general unit vector $\mathbf{r}=(\sin \theta
_{r}\cos \phi _{r},\sin \theta _{r}\sin \phi _{r},\cos \theta _{r})$ is
parameterized by the polar and azimuthal angles\textbf{\ }$\theta _{r}%
\mathbf{,\ }\phi _{r}$ in the coordinate frame with $z$-axis along the
direction of the initial spin-polarization. The two orthogonal states\textbf{%
\ }$|\pm \mathbf{r}\rangle $\textbf{\ }are known as spin coherent states of
north- and south- pole gauges.{\textsuperscript{\cite{27,28,36}}} The
eigenstate product of operators\textbf{\ }$\hat{\sigma}\cdot \mathbf{a}$%
\textbf{\ }and $\hat{\sigma}\cdot \mathbf{b}$\textbf{\ }forms\textbf{\ }an
outcome-independent vector base for measuring two spins respectively along
the\textbf{\ }$\mathbf{a}$, $\mathbf{b}$ directions. We label the four base
vectors as

\begin{equation}
\left\vert 1\right\rangle =\left\vert +\mathbf{a},+\mathbf{b}\right\rangle
,\left\vert 2\right\rangle =\left\vert +\mathbf{a},-\mathbf{b}\right\rangle
,\left\vert 3\right\rangle =\left\vert -\mathbf{a},+\mathbf{b}\right\rangle
,\left\vert 4\right\rangle =\left\vert -\mathbf{a},-\mathbf{b}\right\rangle
\label{vec}
\end{equation}%
for the sake of simplicity. The measurement correlation operator is denoted
by\textbf{\ }%
\begin{equation*}
\hat{\Omega}(ab)=(\hat{\sigma}\cdot \mathbf{a})(\hat{\sigma}\cdot \mathbf{b})%
\mathbf{.}
\end{equation*}%
The correlation probability{\textsuperscript{\cite{25,26}}} is obtained as%
\begin{equation}
P(a,b)=Tr[\hat{\Omega}(a,b)\hat{\rho}]  \label{4}
\end{equation}%
which can be also separated to local and non-local parts%
\begin{equation*}
P(a,b)=P_{lc}(a,b)+P_{nlc}(a,b)
\end{equation*}%
with%
\begin{equation*}
P_{lc}(a,b)=Tr[\hat{\Omega}(a,b)\hat{\rho}_{lc}]
\end{equation*}%
and 
\begin{equation*}
P_{nlc}(a,b)=Tr[\hat{\Omega}(a,b)\hat{\rho}_{nlc}].
\end{equation*}%
In terms of the outcome-independent base vectors given by Eq.(\ref{vec}) we
derive the well known measurement correlation for the local realist model 
\begin{equation*}
P_{lc}(a,b)=\rho _{11}^{lc}-\rho _{22}^{lc}-\rho _{33}^{lc}+\rho
_{44}^{lc}=-\cos \theta _{a}\cos \theta _{b},
\end{equation*}%
which independent of the state parameters $\xi $, $\eta $ is valid for
arbitrary normalized entangled states Eq.(\ref{1}). The BI and CHSH
inequalities are recovered with this correlation.{\textsuperscript{%
\cite{25,26}}} The non-local part found as%
\begin{equation*}
P_{nlc}(a,b)=2\sin \xi \cos \xi \sin \theta _{a}\sin \theta _{b}\cos \left(
\phi _{a}-\phi _{b}+2\eta \right)
\end{equation*}%
however depends on the specific states. The violation of BI is seen to be a
direct result of the non-local correlation. Particularly when the initial
entangled state is the two-spin singlet%
\begin{equation*}
|\psi _{s}\rangle =\frac{1}{\sqrt{2}}\left( |+,-\rangle -|-,+\rangle \right)
,
\end{equation*}%
with the state parameters $\xi =(3\pi /4)\func{mod}2\pi $ and $\eta =0\func{%
mod}2\pi $, the total correlation $P(a,b)$ becomes a scaler product of the
two unit vectors 
\begin{equation*}
P(a,b)=-\mathbf{a}\cdot \mathbf{b,}
\end{equation*}%
from which the BIs are violated. A maximum violation value for the CHSH
correlation is found as 
\begin{equation*}
P_{CHSH}^{\max }=\left\vert P(a,b)+P(a,c)+P(d,b)-P(d,c)\right\vert =2\sqrt{2}%
.
\end{equation*}%
\ 

\part{\protect\large 2.2 Particle-number correlation probability}

The particle-number correlation probability is considered in the Wigner
formalism instead of the spin measuring outcome correlation. For example 
\begin{equation}
N(+a,+b)=|\langle +a,+b|\psi \rangle |^{2}=\langle +a,+b|\hat{\rho}%
|+a,+b\rangle =\rho _{11}  \label{N}
\end{equation}%
describes the particle number correlation probability for two positive-spin
particles respectively along $\mathbf{a}$, $\mathbf{b}$ directions. The WI
can be recovered in terms of the quantum probability statistics with the
particle number correlation $N_{lc}\left( +a,+b\right) =\rho _{11}^{lc}$ of
local realist model. Correspondingly three more correlations are related to
the elements of density operator by 
\begin{equation}
N(+a,-b)=\rho _{22},N(-a,+b)=\rho _{33},N(-a,-b)=\rho _{44},  \label{N1}
\end{equation}%
which are all positive quantities different from the spin measuring-outcome
correlations.

\part{\protect\Large 3. Wigner inequality and upper-bound of violation for
two-spin entangled state with antiparallel spin-polarization}

We consider the two-spin entangled state with antiparallel spin-polarization
in Eq.(\ref{1}). The WI is given by{\textsuperscript{\cite{25,29}}}%
\begin{equation}
N_{lc}\left( +a,+b\right) \leq N_{lc}\left( +a,+c\right) +N_{lc}\left(
+c,+b\right)  \label{wig}
\end{equation}%
in which only the number probability of positive spin is assumed to be
detected along all three directions ($\mathbf{a}$, $\mathbf{b}$, and $%
\mathbf{c}$) for both particles. From the viewpoint of symmetry we extend
the original inequality Eq.(\ref{wig}) to that including also the number
probability of the negative spin. The particle number
correlation-probability along two directions $\mathbf{a}$, $\mathbf{b}$ can
be obtained in terms of Eq.(\ref{N}) and Eqs.(\ref{N1}) 
\begin{equation*}
N(\pm a,\pm b)=N_{lc}(\pm a,\pm b)+N_{nlc}(\pm a,\pm b).
\end{equation*}%
for the arbitrary entangled state Eq.(\ref{1}) with two particles of both
positive and negative spins respectively. Following the same procedure in
the above section\textbf{\ }we have $N_{lc}\left( +a,+b\right) =\rho
_{11}^{lc}$, $N_{lc}\left( -a,-b\right) =\rho _{44}^{lc}$ \ It is a simple
algebra to find%
\begin{equation}
N_{lc}\left( +a,+b\right) =\sin ^{2}\xi \cos ^{2}\frac{\theta _{a}}{2}\sin
^{2}\frac{\theta _{b}}{2}+\cos ^{2}\xi \sin ^{2}\frac{\theta _{a}}{2}\cos
^{2}\frac{\theta _{b}}{2}  \label{ab}
\end{equation}%
\ \ \ \ \ \ \ \ \ \ \ \ \ \ \ \ \ \ \ \ \ \ \ \ \ \ \ \ \ \ \ \ \ \ \ \ \ \
\ \ \ \ \ \ \ \ \ \ \ \ \ \ \ \ \ \ \ \ \ \ \ \ \ \ \ \ \ \ \ \ \ \ \ \ \ \
\ \ \ \ \ \ \ \ \ \ \ \ \ \ \ \ \ \ \ \ \ \ \ \ \ \ \ \ \ \ \ \ \ \ \ \ \ \
\ \ \ \ \ \ \ \ \ \ \ \ \ \ \ \ \ \ \ \ \ \ \ \ \ \ \ \ \ \ \ \ \ \ \ \ \ \
\ \ \ \ \ \ \ \ \ \ \ \ \ \ \ \ \ \ \ \ \ \ \ \ \ \ \ \ \ \ \ \ \ \ \ \ \ \
\ \ \ \ \ \ \ \ \ \ \ \ \ \ \ \ \ \ \ \ \ \ \ \ \ \ \ \ \ \ \ \ \ \ \ \ \ \
\ \ \ \ \ \ \ \ \ \ \ \ \ \ \ \ \ \ \ \ \ \ \ \ \ \ \ \ \ \ \ \ \ \ \ \ \ \
\ \ \ \ \ \ \ \ \ \ \ \ \ \ \ \ \ \ \ \ \ \ \ \ \ \ \ \ \ \ \ \ \ \ \ \ \ \
\ \ \ \ \ \ \ \ \ \ \ \ \ \ \ \ \ \ \ \ \ \ \ \ \ \ \ \ \ \ \ \ \ \ \ \ \ \
\ \ \ \ \ \ \ \ \ \ \ \ \ \ \ \ \ \ \ \ \ \ \ \ \ \ \ \ \ \ \ \ \ \ \ \ \ \
\ \ \ \ \ \ \ \ \ \ \ \ \ \ \ \ \ \ \ \ \ \ \ \ \ \ \ \ \ \ \ \ \ \ \ \ \ \
\ \ \ \ \ \ \ \ \ \ \ \ \ \ \ \ \ \ \ \ \ \ \ \ \ \ \ \ \ \ \ \ \ \ \ \ \ \
\ \ \ \ \ \ \ \ \ \ \ \ \ \ \ \ \ \ \ \ \ \ \ \ \ \ \ \ \ \ \ \ \ \ \ \ \ \
\ \ \ \ \ \ \ \ \ \ \ \ \ \ \ \ \ \ \ \ \ \ \ \ \ \ \ \ \ \ \ \ \ \ \ \ \ \
\ \ \ \ \ \ \ \ \ \ \ \ \ \ \ \ \ \ \ \ \ \ \ \ \ \ \ \ \ \ \ \ \ \ \ \ \ \
\ \ \ \ \ \ \ \ \ \ \ \ \ \ \ \ \ \ \ \ \ \ \ \ \ \ \ \ \ \ \ \ \ \ \ \ \ \
\ \ \ \ \ \ \ \ \ \ \ \ \ \ \ \ \ \ \ \ \ \ \ \ \ \ \ \ \ \ \ \ \ \ \ \ \ \
\ \ \ \ \ \ \ \ \ \ \ \ \ \ \ \ \ \ \ \ \ \ \ \ \ \ \ \ \ \ \ \ \ \ \ \ \ \
\ \ \ \ \ \ \ \ \ \ \ \ \ \ \ \ \ \ \ \ \ \ \ \ \ \ \ \ \ \ \ \ \ \ \ \ \ \
\ \ \ \ \ \ \ \ \ \ \ \ \ \ \ \ \ \ \ \ \ \ \ \ \ \ \ \ \ \ \ \ \ \ \ \ \ \
\ \ \ \ \ \ \ \ \ \ \ \ \ \ \ \ \ \ \ \ \ \ \ \ \ \ \ \ \ \ \ \ \ \ \ \ \ \
\ \ \ \ \ \ \ \ \ \ \ \ \ \ \ \ \ \ \ \ \ \ \ \ \ \ \ \ \ \ \ \ \ \ \ \ \ \
\ \ \ \ \ \ \ \ \ \ \ \ \ \ \ \ \ \ \ \ \ \ \ \ \ \ \ \ \ \ \ \ \ \ \ \ \ \
\ \ \ \ \ \ \ \ \ \ \ \ \ \ \ \ \ \ \ \ \ \ \ \ \ \ \ \ \ \ \ \ \ \ \ \ \ \
\ \ \ \ \ \ \ \ \ \ \ \ \ \ \ \ \ \ \ \ \ \ \ \ \ \ \ \ \ \ \ \ \ \ \ \ \ \
\ \ \ \ \ \ \ \ \ \ \ \ \ \ \ \ \ \ \ \ \ \ \ \ \ \ \ \ \ \ \ \ \ \ \ \ \ \
\ \ \ \ \ \ \ and 
\begin{equation}
N_{lc}\left( -a,-b\right) =\sin ^{2}\xi \sin ^{2}\frac{\theta _{a}}{2}\cos
^{2}\frac{\theta _{b}}{2}+\cos ^{2}\xi \cos ^{2}\frac{\theta _{a}}{2}\sin
^{2}\frac{\theta _{b}}{2}  \label{-a-b}
\end{equation}%
which depend on the state parameter $\xi $. The correlation probabilities of
two-direction measurements are different for the positive and negative spin
particles. However, we are going to show an interesting fact that the WI
itself is independent of the state parameter $\xi $. It is also the same no
matter whether the positive or negative spin particles are measured.

To have a quantitative bound for the violation of WI, we following CHSH
define a correlation probability for the three-direction measurement 
\begin{equation*}
W_{lc}=N_{lc}\left( \pm a,\pm b\right) -N_{lc}\left( \pm a,\pm c\right)
-N_{lc}\left( \pm c,\pm b\right) .
\end{equation*}%
Then the original form of WI is equivalent to 
\begin{equation*}
0\geq W_{lc}.
\end{equation*}%
Substitution of the corresponding correlations $N_{lc}\left( \pm a,\pm
b\right) $, $N_{lc}\left( \pm a,\pm c\right) $ and $N_{lc}\left( \pm c,\pm
b\right) $ into $W_{lc}$ yields\ 
\begin{equation}
W_{lc}=-\left( \cos ^{2}\frac{\theta _{a}}{2}-\cos ^{2}\frac{\theta _{c}}{2}%
\right) \cos ^{2}\frac{\theta _{b}}{2}-\cos ^{2}\frac{\theta _{c}}{2}\sin
^{2}\frac{\theta _{a}}{2},  \label{lc}
\end{equation}%
for both positive and negative spins. From Eq.(\ref{lc}) we can verify after
a simple algebra the inequality that 
\begin{equation*}
0\geq -\sin ^{2}\frac{\theta _{c}}{2}\cos ^{2}\frac{\theta _{a}}{2}\geq
W_{lc}.
\end{equation*}%
Therefore the original form{\textsuperscript{\cite{29}}} of WI is satisfied
not only for detection of the positive but also the negative spin
particle-numbers.

The non-local part of correlation is $N_{nlc}\left( +a,+b\right) =$ $\rho
_{11}^{nlc},$ $N_{nlc}\left( -a,-b\right) =\rho _{44}^{nlc}$. Since $\rho
_{11}^{nlc}=\rho _{44}^{nlc}$, we have 
\begin{equation}
N_{nlc}\left( \pm a,\pm b\right) =\frac{1}{2}\sin \xi \cos \xi \sin \theta
_{a}\sin \theta _{b}\cos \left( \phi _{a}-\phi _{b}+2\eta \right) ,
\label{nlcanti}
\end{equation}%
which depends on the state parameters $\xi $, $\eta $. Including the
non-local part the three-direction correlation probability becomes%
\begin{equation*}
W=N\left( \pm a,\pm b\right) -N\left( \pm a,\pm c\right) -N\left( \pm c,\pm
b\right) =W_{lc}+W_{nlc},
\end{equation*}%
where the non-local part%
\begin{align}
W_{nlc}& =\frac{1}{4}\sin (2\xi )[\sin \theta _{a}\sin \theta _{b}\cos
\left( \phi _{a}-\phi _{b}+2\eta \right)  \label{nlc} \\
& -\sin \theta _{a}\sin \theta _{c}\cos \left( \phi _{a}-\phi _{c}+2\eta
\right)  \notag \\
& -\sin \theta _{c}\sin \theta _{b}\cos \left( \phi _{c}-\phi _{b}+2\eta
\right) ],  \notag
\end{align}%
depends also on the state parameters.

We now analyze the violation of WI by the non-local part of correlation $%
W_{nlc}$. Since the polar angles are restricted by $0\leq \theta \lessdot
\pi $, the non-local probability of Eq.(\ref{nlc}) obeys the following
inequality%
\begin{equation*}
W_{nlc}\leq \frac{1}{4}\left( \sin \theta _{a}\sin \theta _{b}+\sin \theta
_{a}\sin \theta _{c}+\sin \theta _{c}\sin \theta _{b}\right) .
\end{equation*}%
On other hand the local part Eq.(\ref{lc}) can be rewritten as%
\begin{equation}
W_{lc}=\frac{1}{4}(-1-\cos \theta _{a}\cos \theta _{b}+\cos \theta _{c}\cos
\theta _{b}+\cos \theta _{a}\cos \theta _{c}).  \label{lc2}
\end{equation}%
Adding them together we obtain an inequality obeyed by the Wigner
correlation that%
\begin{equation}
W\leq F(\theta _{a}\text{, }\theta _{b}\text{, }\theta _{c}),  \label{f}
\end{equation}%
where%
\begin{equation*}
F(\theta _{a}\text{, }\theta _{b}\text{, }\theta _{c})=\frac{1}{4}[-1-\cos
\left( \theta _{a}+\theta _{b}\right) +\cos \left( \theta _{c}-\theta
_{b}\right) +\cos \left( \theta _{a}-\theta _{c}\right) ].
\end{equation*}%
Since the function $F(\theta _{a}$, $\theta _{b}$, $\theta _{c})$ can be
greater than zero, the WI is then violated. It is easy to prove that 
\begin{equation*}
F(\theta _{a}\text{, }\theta _{b}\text{, }\theta _{c})\leq \frac{1}{2}.
\end{equation*}%
We thus derive a maximum violation bound 
\begin{equation}
W_{\max }=\frac{1}{2},  \label{max}
\end{equation}%
which is universal for arbitrary entangled state and any three-direction
measurements.

As a matter of fact, for the state-parameter angles\textbf{\ }$\xi =\pi /4%
\func{mod}2\pi $\textbf{\ }and\textbf{\ }$\eta =0\func{mod}2\pi $, the
two-spin entangled state becomes the two-spin triplet with the vanishing
magnetic-eigenvalue $m=0$ that 
\begin{equation*}
|\psi _{t}\rangle =\frac{1}{\sqrt{2}}\left( |+,-\rangle +|-,+\rangle \right)
.
\end{equation*}%
Corresponding non-local part of measuring outcome probability becomes%
\begin{align*}
W_{nlc}& =\frac{1}{4}[\sin \theta _{a}\sin \theta _{b}\cos \left( \phi
_{a}-\phi _{b}\right) -\sin \theta _{a}\sin \theta _{c}\cos \left( \phi
_{a}-\phi _{c}\right) \\
& -\sin \theta _{c}\sin \theta _{b}\cos \left( \phi _{c}-\phi _{b}\right) ].
\end{align*}%
If $\theta _{a}=\theta _{c}=\theta _{b}=\pi /2$, $\phi _{a}=\phi _{b}=\pi $, 
$\phi _{c}=0$, namely $\mathbf{a},\mathbf{b},\mathbf{c}$ are perpendicular
to the original spin polarization with $\mathbf{a},\mathbf{b}$ along $-x$%
-directions $\mathbf{c}$ along $x$-direction, the Wigner correlation
reaches, in this case, the maximum violation bound, $W_{\max }=1/2$.

\part{\protect\Large 4. Parallel spin polarization}

We now consider the two-spin entangled state with parallel polarization{%
\textsuperscript{\cite{26}}} 
\begin{equation*}
\left\vert \psi \right\rangle _{pl}=c_{1}\left\vert +,+\right\rangle
+c_{2}\left\vert -,-\right\rangle
\end{equation*}%
in which two arbitrary coefficients are parameterized as before. The density
operator $\hat{\rho}_{pl}$ is also separated to local part%
\begin{equation*}
\hat{\rho}_{lc}^{pl}=\sin ^{2}\xi |+,+\rangle \left\langle +,+\right\vert
+\cos ^{2}\xi |-,-\rangle \left\langle -,-\right\vert
\end{equation*}%
and the non-local part%
\begin{equation*}
\hat{\rho}_{nlc}^{pl}=\sin \xi \cos \xi \left( e^{2i\eta }|+,+\rangle
\left\langle -,-\right\vert +e^{-2i\eta }|-,-\rangle \left\langle
+,+\right\vert \right) .
\end{equation*}%
We find that particles with opposite spins have to be detected respectively
for the two directions, namely

\begin{equation*}
N_{pl}(\pm a,\mp b)=N_{lc}^{pl}(\pm a,\mp b)+N_{nlc}^{pl}(\pm a,\mp b).
\end{equation*}%
The local part of particle-number correlation with positive-spin particle
detected in $a$-direction and negative-spin in $b$-direction is evaluated as 
\begin{equation*}
N_{lc}^{pl}\left( +a,-b\right) =\left( \rho _{lc}^{pl}\right)
_{22}=N_{lc}(+a,+b),
\end{equation*}%
which equals exactly the correlation probability $N_{lc}(+a,+b)$ in Eq.(\ref%
{ab}) for the antiparallel case. While the local correlation-probability for 
$a$-direction negative and $b$-direction positive is 
\begin{equation*}
N_{lc}^{pl}\left( -a,+b\right) =\left( \rho _{lc}^{pl}\right)
_{33}=N_{lc}\left( -a,-b\right) ,
\end{equation*}%
which equals $N_{lc}\left( -a,-b\right) $ in Eq.(\ref{-a-b}) for the
antiparallel case. Thus, the Wigner correlation probability for the
entangled state with parallel spin-polarizations is the same as the
antiparallel case given in Eq.(\ref{lc})

\begin{equation*}
N_{lc}^{pl}\left( \pm a,\mp b\right) -N_{lc}^{pl}\left( \pm a,\mp c\right)
-N_{lc}^{pl}\left( \pm c,\mp b\right) =W_{lc}\leq 0.
\end{equation*}%
The validity of WI for parallel spin-polarizations is also verified in
Appendix by means of classical statistics following the original work of
Wigner.{\textsuperscript{\cite{29,30}}}

The non-local parts, which result in the violation of WI, are evaluated from
the density operator elements of entangled state with parallel
spin-polarization, such that%
\begin{equation*}
N_{nlc}^{pl}\left( +a,-b\right) =(\rho _{nlc}^{pl})_{22},\quad
N_{nlc}^{pl}\left( -a,+b\right) =(\rho _{nlc}^{pl})_{33}.
\end{equation*}%
We find that the interchange of detecting positive and negative spin
particles in the two directions gives rise to the same result that%
\begin{equation*}
N_{nlc}^{pl}\left( \pm a,\mp b\right) =-\frac{1}{2}\sin \xi \cos \xi \sin
\theta _{a}\sin \theta _{b}\cos \left( \phi _{a}+\phi _{b}+2\eta \right) .
\end{equation*}%
The total non-local part for three-direction measurements is seen to be%
\begin{align}
W_{nlc}^{pl}& =-\frac{1}{4}\sin \left( 2\xi \right) [\sin \theta _{a}\sin
\theta _{b}\cos \left( \phi _{a}+\phi _{b}+2\eta \right)  \label{nlc2} \\
& -\sin \theta _{a}\sin \theta _{c}\cos \left( \phi _{a}+\phi _{c}+2\eta
\right)  \notag \\
& -\sin \theta _{c}\sin \theta _{b}\cos \left( \phi _{c}+\phi _{b}+2\eta
\right) ].  \notag
\end{align}%
Following the same procedure of analyses as in the antiparallel case we
again have the maximum violation bound $W_{\max }=1/2$. Thus we conclude
that WI and its violation are universal for arbitrary two-spin entangled
states with both antiparallel and parallel spin-polarizations. As an example
we consider a particular entangled state 
\begin{equation*}
|\psi _{pl}\rangle =\frac{1}{\sqrt{2}}\left( |+,+\rangle +|-,-\rangle
\right) ,
\end{equation*}%
resulted by the parameter angles\textbf{\ }$\xi =\pi /4\func{mod}2\pi $%
\textbf{\ }and\textbf{\ }$\eta =0\func{mod}2\pi $. Corresponding non-local
part of correlation probability in Eq.(\ref{nlc2}) becomes

\begin{align*}
W_{nlc}^{pl}& =-\frac{1}{4}[\sin \theta _{a}\sin \theta _{b}\cos \left( \phi
_{a}+\phi _{b}\right) -\sin \theta _{a}\sin \theta _{c}\cos \left( \phi
_{a}+\phi _{c}\right) \\
& -\sin \theta _{c}\sin \theta _{b}\cos \left( \phi _{c}+\phi _{b}\right) ].
\end{align*}%
The maximum violation $W_{\max }=1/2$ can be approached when $\theta
_{a}=\theta _{c}=\theta _{b}=\pi /2$, $\phi _{a}=\phi _{b}=\pi /2$, $\phi
_{c}=3\pi /2$, namely the three measuring directions are colinear with $%
\mathbf{a},\mathbf{b}$ along $y$-direction and $\mathbf{c}$ along $-y$%
-direction in the chosen coordinate frame with initial spin-polarization in $%
z$-axis.

\part{\protect\Large 5. Conclusion and Discussion}

By means of the spin coherent-state quantum probability statistics, the
original WI is extended to arbitrary two-spin entangled states with
antiparallel and parallel spin polarizations. For the antiparallel case both
positive or both negative spin particles ought to be detected respectively
in three directions. While opposite spin measurements are necessary for the
parallel case. The Wigner correlation $W_{lc}$ is always less than or at
most equal to zero by the local realist-theory. A maximum violation $W_{\max
}=1/2$ is found for arbitrary two-spin entangled states with both parallel
and antiparallel spin-polarizations. The measured violation-value depends on
specific state, which is parameterized by parameters $\xi $ and $\eta $, and
also on the measuring directions. The positive and negative spin particles
can be detected by the Stern-Gerlach experiment with a gradient
magnetic-field. A loophole-free experimental verification of the violation
of CHSH inequality was reported recently by means of electronic spin
associated with a single nitrogen-vacancy defect centre in a diamond chip.{%
\textsuperscript{\cite{37}} The experimental verification of the WI
violation is also expected.}

We conclude that the WI is equally convenient for the experimental
verification of its violation. It may be more suitable for any bipartite
systems besides the two spins since only particle number probabilities are
needed for detection but not the spin variables. Although our formalism is
based on two-spin entangled state, the result can be used for the two-photon
entangled states with perpendicular polarizations.{\textsuperscript{%
\cite{38}}}

\part{\protect\large Acknowledge}

This work was supported in part by National Natural Science Foundation of
China, under Grants No. 11275118, U1330201.

\part{\protect\large Appendix}

The extended WI $N_{lc}\left( \pm a,\mp b\right) \leq N_{lc}\left( \pm a,\mp
c\right) +N_{lc}\left( \pm c,\mp b\right) $ can be proved in terms of
classical statistics for\ two-spin entangled state with parallel
polarizations. Eight independent particle-number probabilities are denoted by%
{\textsuperscript{\cite{30}}}

\ \ \ \ \ \ \ \ \ \ \ \ \ \ \ \ \ \ \ \ \ \ \ \ \ \ \ \ \ \ \ \ $\ \ \ \ \ \
\ \ \ \ \ \ \ \ \ \ \ \ \ \ \ \ \ \ \ \ \ \ \ \ \ \ \ \ \ \ \ \ $%
\begin{eqnarray*}
&&\ \text{Table.\ Spin-correlation\ Measurements} \\
&&%
\begin{tabular}{ccc}
\hline
population & particle1 & particle2 \\ \hline
$N_{1}$ & $\left( +a,+b,+c\right) $ & $\left( +a,+b,+c\right) $ \\ \hline
$N_{2}$ & $\left( +a,+b,-c\right) $ & $\left( +a,+b,-c\right) $ \\ \hline
$N_{3}$ & $\left( +a,-b,+c\right) $ & $\left( +a,-b,+c\right) $ \\ \hline
$N_{4}$ & $\left( +a,-b,-c\right) $ & $\left( +a,-b,-c\right) $ \\ \hline
$N_{5}$ & $\left( -a,+b,+c\right) $ & $\left( -a,+b,+c\right) $ \\ \hline
$N_{6}$ & $\left( -a,+b,-c\right) $ & $\left( -a,+b,-c\right) $ \\ \hline
$N_{7}$ & $\left( -a,-b,+c\right) $ & $\left( -a,-b,+c\right) $ \\ \hline
$N_{8}$ & $\left( -a,-b,-c\right) $ & $\left( -a,-b,-c\right) $ \\ \hline
\end{tabular}%
\end{eqnarray*}%
for measurement of two spin-particles along unit-vector directions $\mathbf{a%
}$, $\mathbf{b}$ and $\mathbf{c}$, respectively. The measuring outcome
correlation probabilities among the three directions are represented in
terms of the population probabilities such that

$\ \ \ \ \ \ \ \ \ \ \ \ \ \ \ \ \ \ \ \ \ $%
\begin{equation*}
N_{lc}\left( +a,-b\right) =\frac{\left( N_{3}+N_{4}\right) }{%
\sum_{i}^{8}N_{i}},
\end{equation*}%
\begin{equation*}
N_{lc}\left( +a,-c\right) =\frac{\left( N_{2}+N_{4}\right) }{%
\sum_{i}^{8}N_{i}},
\end{equation*}%
and%
\begin{equation*}
N_{lc}\left( +c,-b\right) =\frac{\left( N_{3}+N_{7}\right) }{%
\sum_{i}^{8}N_{i}}.
\end{equation*}%
Since 
\begin{equation*}
N_{3}+N_{4}\leq \left( N_{2}+N_{4}\right) +\left( N_{3}+N_{7}\right) ,
\end{equation*}%
thus we have the WI 
\begin{equation*}
N_{lc}\left( +a,-b\right) \leq N_{lc}\left( +a,-c\right) +N_{lc}\left(
+c,-b\right) .
\end{equation*}%
Alternatively interchange of the measuring positive and negative
spin-particles leads to%
\begin{equation*}
N_{lc}\left( -a,+b\right) =\frac{\left( N_{5}+N_{6}\right) }{%
\sum_{i}^{8}N_{i}},
\end{equation*}%
\begin{equation*}
N_{lc}\left( -a,+c\right) =\frac{\left( N_{5}+N_{7}\right) }{%
\sum_{i}^{8}N_{i}},
\end{equation*}%
and%
\begin{equation*}
N_{lc}\left( -c,+b\right) =\frac{\left( N_{2}+N_{6}\right) }{%
\sum_{i}^{8}N_{i}}.
\end{equation*}%
Since%
\begin{equation*}
N_{5}+N_{6}\leq \left( N_{5}+N_{7}\right) +\left( N_{2}+N_{6}\right) ,
\end{equation*}%
we have again the WI%
\begin{equation*}
N_{lc}\left( -a,+b\right) \leq N_{lc}\left( -a,+c\right) +N_{lc}\left(
-c,+b\right) .
\end{equation*}

\bigskip

\part


{\protect\large References}

\begin{thebibliography}{99}
\bibitem{1} {\small S. Groblacher, T. Paterek, R. Kaltenbaek, C. Brukner, M.
Z. Ukowski, M. Aspelmeyer, and A. Zeilinger, \textit{Nature} \textbf{446}
(2007) 871.}

\bibitem{2} {\small H. Buhrman, R. Cleve, S. Massar, and R. de Wolf, \textit{%
Rev. Mod. Phys}. \textbf{82} (2010) 665.}

\bibitem{3} {\small C. Branciard, A. Ling, N. Gisin, C. Kurtsiefer, A.
Lamas-Linares, and V. Scarani, \textit{Phys. Rev. Lett.} \textbf{99} (2007)
210407.}

\bibitem{4} {\small M. D. Eisaman, E. A. Goldschmidt, J. Chen, J. Fan, and
A. Migdall, \textit{Phys. Rev. A} \textbf{77} (2008) 032339.}

\bibitem{5} {\small M. Paternostro and H. Jeong, \textit{Phys. Rev. A }%
\textbf{81} (2010) 032115.}

\bibitem{6} {\small C.-W. Lee, M. Paternostro, and H. Jeong, \textit{Phys.
Rev. A} \textbf{83} (2011) 022102.}

\bibitem{7} {\small S. Pironio \textit{et al.}, \textit{Nature} \textbf{464}
(2010) 1021.}

\bibitem{8} {\small D. Bohm, \textit{Phys. Rev.} \textbf{108} (1957) 1070.}

\bibitem{9} {\small J. S. Bell, \textit{Physics} \textbf{1} (1964) 195.}

\bibitem{10} {\small L. F. Wei, Y. X. Liu, and F. Nori, \textit{Phys. Rev. B 
}\textbf{72} (2005) 104516; M. Ansmann, \textit{et al.}, \textit{Nature} 
\textbf{461} (2009) 504.}

\bibitem{11} {\small S. Groblacher, T. Paterek, R. Kaltenbaek, C. Brukner,
M. Z. Ukowski, M. Aspelmeyer, and A. Zeilinger, \textit{Nature} \textbf{446}
(2007) 871.}

\bibitem{12} {\small H. Buhrman, R. Cleve, S. Massar, and R. de Wolf, 
\textit{Rev. Mod. Phys.} \textbf{82} (2010) 665.}

\bibitem{13} {\small L. Allen, M. W. Beijersbergen, R. J. C. Spreeuw, and J.
P. Woerdman, \textit{Phys. Rev. A} \textbf{45} (1992) 8185.}

\bibitem{14} {\small A. Cabello and F. Sciarrino,\textit{\ Phys. Rev. X} 
\textbf{2} (2012) 021010.}

\bibitem{15} {\small G.-F. Zhang, \textit{et al}., \textit{Annals of Physics}
\textbf{326}(10) (2011) 2694-2701.}

\bibitem{16} {\small A. Aspect, \textit{Nature} \textbf{398} (1999) 189.}

\bibitem{17} {\small A. C. Dada, J. Leach, G. S. Buller, M. J. Padgett, E.
Andersson, \textit{Nature Phys.} \textbf{7} (2011) 677.}

\bibitem{18} {\small W. Tittel, J. Brendel, B. Gisin, T. Herzog, H. Zbinden,
and N. Gisin, \textit{Phys. Rev. A} \textbf{57} (1998) 3229; W. Tittel, J.
Brendel, H. Zbinden, and N. Gisin, \textit{Phys. Rev. Lett.} \textbf{81}
(1998) 3563.}

\bibitem{19} {\small G. Weihs, T. Jennewein, C. Simon, H. Weinfurter, and A.
Zeilinger,\textit{\ Phys. Rev. Lett.} \textbf{81} (1998) 5039.}

\bibitem{20} {\small M. Rowe,\textit{\ et al.}, \textit{Nature} \textbf{409}
(2001) 791.}

\bibitem{21} {\small H. Sakai,\textit{\ et al.}, \textit{Phys. Rev. Lett.} 
\textbf{97} (2006) 150405.}

\bibitem{22} {\small D. Kaszlikowski, P. Gnaci\'{n}ski, M. \.{Z}ukowski, W.
Miklaszewski, and A. Zeilinger, \textit{Phys. Rev. Lett.} \textbf{85} (2000)
4418.}

\bibitem{23} {\small J. R. Torgerson, D. Branning, C. H. Monken, L. Mandel, 
\textit{Phys. Lett. A} \textbf{204}(5-6) (1995) 323-328.}

\bibitem{24} {\small J. F. Clauser, M. A. Horne, A. Shimony and R. A. Holt, 
\textit{Phys. Rev. Lett.} \textbf{23} (1969) 880.}

\bibitem{25} {\small Z. Song, J.-Q. Liang, and L.-F. Wei,\ \textit{Mod.
Phys. Lett. B} \textbf{28}(1) (2014) 1450004.}

\bibitem{26} {\small H. Zhang, J. Wang, Z. Song, J.-Q.Liang, L.-F. Wei, 
\textit{Mod. Phys. Lett. B} \textbf{31}(4) (2017) 1750032.}

\bibitem{27} {\small J.-Q. Liang and L. F. Wei, \textit{New Advances in
Quantum Physics}, (Science Press, Beijing, 2011).}

\bibitem{28} {\small X.-M. Bai, C.-P. Gao, J.-Q. Li, J.-Q. Liang, \textit{%
Optics express}, \textbf{25} (2017) 17051-17065.}

\bibitem{29} {\small E. P. Wigner, \textit{Am. J. Phys.} \textbf{38} (1970)
1005.}

\bibitem{30} {\small J. J. Sakurai, S. F. Tuan, and E. D. Commins, Modern
Quantum Mechanics revised edition, \textit{Am. J. Phys.} \textbf{63} (1995)
93.}

\bibitem{31} {\small N. Gisin, \textit{Phys. Lett. A} \textbf{154} (1991)
201.}

\bibitem{32} {\small N. Gisin and A. Peres,\textit{\ Phys. Lett. A} \textbf{%
162} (1992) 15.}

\bibitem{33} {\small F. A. Bovino, I. P. Degiovanni, \textit{Phys. Rev. A} 
\textbf{77} (2008) 052110.}

\bibitem{34} {\small D. Home, D. Saha, S. Das,\textit{\ Phys. Rev. A }%
\textbf{91}(1) (2015) 012102.}

\bibitem{35} {\small D. Das, S. Datta, S. Goswami, A. S. Majumdar, D. Home,%
\textit{\ Phys. Lett. A} \textbf{381}(39) (2017) 3396-3404.}

\bibitem{36} {\small X. -Q. Zhao, N. Liu, J. -Q. Liang,\ \textit{Phys. Rev. A%
} \textbf{90} (2014) 023622.}

\bibitem{37} {\small B. Hensen, \textit{et al.},\textit{\ Nature} \textbf{526%
} (2015) 682;\ \textit{Scientific Reports} \textbf{6} (2016) 30289.}

\bibitem{38} {\small J. Yin \textit{et al.}, \textit{Science} \textbf{356}
(2017) 1140--1144.}
\end{thebibliography}
\end{document}